\documentclass[prb,twocolumn,superscriptaddress,preprintnumbers,a4paper,amsmath,amssymb,showpacs,floatfix]{revtex4-1}
\usepackage{graphicx}
\usepackage{graphics}
\usepackage{bm}
\usepackage{dcolumn}

\newcommand{\urrs}{U(Ru$_{1-x}$Rh$_{x}$)$_{2}$Si$_{2}$}
\newcommand{\urxrs}{U(Ru$_{0.92}$Rh$_{0.08}$)$_{2}$Si$_{2}$}
\newcommand{\urs}{URu$_{2}$Si$_{2}$}
\newcommand{\tcs}{ThCr$_{2}$Si$_{2}$}

\begin{document}

\title{Electronic properties of a heavy-fermion \urxrs~single crystal}

\author{K.~Proke\v{s}}
\email{prokes@helmholtz-berlin.de}
\affiliation{Helmholtz-Zentrum
Berlin f\"{u}r Materialien und Energie, Hahn-Meitner Platz 1, 14109 Berlin, Germany}

\author{Y.-K. Huang}
\affiliation{Van der Waals-Zeeman Institute, University of Amsterdam, 1018XE Amsterdam, The Netherlands}

\author{M. Reehuis}
\affiliation{Helmholtz-Zentrum
Berlin f\"{u}r Materialien und Energie, Hahn-Meitner Platz 1, 14109 Berlin, Germany}

\author{B. Klemke}
\affiliation{Helmholtz-Zentrum
Berlin f\"{u}r Materialien und Energie, Hahn-Meitner Platz 1, 14109 Berlin, Germany}

\author{J.-U.~Hoffmann}
\email{prokes@helmholtz-berlin.de}
\affiliation{Helmholtz-Zentrum
Berlin f\"{u}r Materialien und Energie, Hahn-Meitner Platz 1, 14109 Berlin, Germany}

\author{A. Sokolowski}
\affiliation{Helmholtz-Zentrum
Berlin f\"{u}r Materialien und Energie, Hahn-Meitner Platz 1, 14109 Berlin, Germany}

\author{J.A.~Mydosh}
\affiliation{Kamerlingh Onnes Laboratory and Institute Lorentz, Leiden University, 2300 RA Leiden, The Netherlands }

\date{\today}

\begin{abstract}
We report the crystal structure and  highly-anisotropic magnetic, transport and thermal properties of an exceptionally good single crystal of \urxrs~prepared using a modified Czochralski method. Our study, that also includes neutron diffraction results, shows all the heavy-fermion signatures of pristine \urs, however, the superconductivity, hidden order and remanent weak antiferromagnetic orders are absent. Instead, the ground state of the doped system can be classified as a spin liquid that preserves the heavy-fermion character. \urxrs~exhibits a short-range magnetic order distinguished by reflections of a Lorentzian profile at \textbf{\textit{q$_ {III}$}} = ($\frac{1}{2}$ $\frac{1}{2}$ $\frac{1}{2}$) positions that disappear above $\approx$ 15 K. The short-range order seems to be a precursor of a long-range magnetic order that occurs with higher Rh concentration. We indicate that these short-range fluctuations involve, at least partially, inelastic scattering processes.

\end{abstract}

\maketitle

\section{Introduction}
\urs~ is a heavy fermion compound (space group $I4/mmm$), $\gamma$ = 180 mJ K$^{2}$mole$^{-1}$, where superconductivity (SC) below the superconducting transition at T$_{sc}$ = 1.5~K coexists with yet unidentified order parameter that is linked to a remanent antiferromagnetism (AFM) \cite{urrs3,urrs14}, both appearing below T$_{HO}$ = 17.5~K (for a review see  \cite{urrs4,urrs5}). The weak AFM order is characterized by a propagation vector \textbf{\textit{q$_ {I}$}} = (1 0 0) with the ordered dipolar moments pointing along the $c$-axis and very small, (0.01 - 0.03 $\mu_{B}$ \cite{urrs8}). In order to explain the clear anomalies in temperature dependencies of 
 properties, notably the large entropy connected with the transition, an unconventional phase change is postulated\cite{urrs3}. It is generally accepted that this AFM order is not intrinsic but parasitic. Accordingly, this new state is called hidden order (HO) and is one of the most addressed topics in heavy-fermion physics research\cite{urrs5}. Near the HO state, different phases can be induced by external perturbations like pressure, magnetic field or substitution. A moderate pressure converts the HO phase \cite{urrs9} into a static long-range antiferromagnetic (AF) q$_ {I}$ order with U magnetic moments of $\approx$ 0.4 $\mu_{B}$, the so-called large moment antiferromagnetic (LMAF) phase.  Here, the tiny remanent AF order in the pure \urs~is usually called the small moment antiferromagnetic (SMAF) phase. The influence of applied magnetic field or substitutions effects are more complicated. A strong magnetic field is necessary to suppress the HO order and generate final Fermi-liquid behavior via intermediate field-induced phases between 35 and 38 T with \textbf{\textit{q$_ {II}$}} = (0.6 0 0) \cite{urrs10,urrs11,urrs26,urrs33}.

It has been demonstrated that few at.\% substitution of various transition metals (Re, Rh, Fe, Os, Tc) for Ru destroy the SC and induce magnetic states. The doping weakens the HO state and transform it to long-range magnetic order that is for different dopant different (the AF order with q$_ {I}$ = (1 0 0) being the most frequent one)\cite{urrs12,urrs17,urrs20,urrs21,urrs22,urrs23,urrs24,urrs25}. In some cases the crossover between the HO state and dopand-induced long-range magnetic order is smooth (as in the case of Fe doping), in other cases a gap between the two ground states exists (the case of Rh) \cite{urrs12,urrs13} preceded by a tiny concentration region where the LMAF phase seems to inhomogeneously coexist.  Focusing on the case of Rh doping, a shift of the large critical fields of metamagnetic transitions to lower fields is found \cite{urrs12,urrs36,urrs27}. Above ~4~\% of Rh substitution for Ru, no trace of HO state or long-range magnetic order can be discerned \cite{urrs12,urrs13}. However, short-range AF correlations start to develop around \textbf{\textit{q$_ {III}$}} = ($\frac{1}{2}$ $\frac{1}{2}$ $\frac{1}{2}$) that “condense” above ~10~\% of Rh for Ru into long-range AF magnetic order with increasing phase transition temperatures \cite{urrs12}. 

The low-concentration Rh region, where the HO coexists with the parasitic LMAF, has been extensively studied in the past \cite{urrs12,urrs13,urrs27,urrs18,urrs36}. Also the higher Rh doping levels, that exhibit magnetic order with high ordering temperatures, have been inspected to a certain extend \cite{urrs12,urrs19}. However, the gap region where no HO or SMAF or LMAF states are present, has been glimped using one 6 \% Rh crystal by Burlet et al. \cite{urrs12}. This region is most interesting as it should be pure Fermi liquid, free of influences of any long-range order and offers a possibility to disclose the "bare"  behavior of heavy electrons in such lightly doped Rh systems. This has motivated us to prepare a high-quality \urxrs~ single crystal (see the resolution-limited rocking curve through the 1 1 0 nuclear Bragg reflection in Fig.~\ref{fig1} and the grown crystal in the inset of Fig.~\ref{fig2}). As we will show below, it indeed does not exhibit any sign of SMAF order or long-range order associated with \textbf{\textit{q$_ {I}$}} or \textbf{\textit{q$_ {III}$}}, respectively. No signatures of HO or SC down to 0.4 K is observed. However, a heavy-fermion behavior remains intact, and since all the anomalies connected with HO and SC phase are removed, this "bare" heavy-fermion state resembles very strongly properties of the pristine \urs. Nevertheless, in contrast to \urs, it exhibits short-range order (SRO) at \textbf{\textit{q$_ {III}$}}.  The SRO signal disappears at temperatures comparable to T$_{HO}$. Our findings suggest that the heavy-fermion behavior is common to all the lightly doped \urs~systems and that the HO transition is a result of coherence phenomenon within the heavy-fermion liquid.

\begin{figure}
\includegraphics*[scale=0.3]{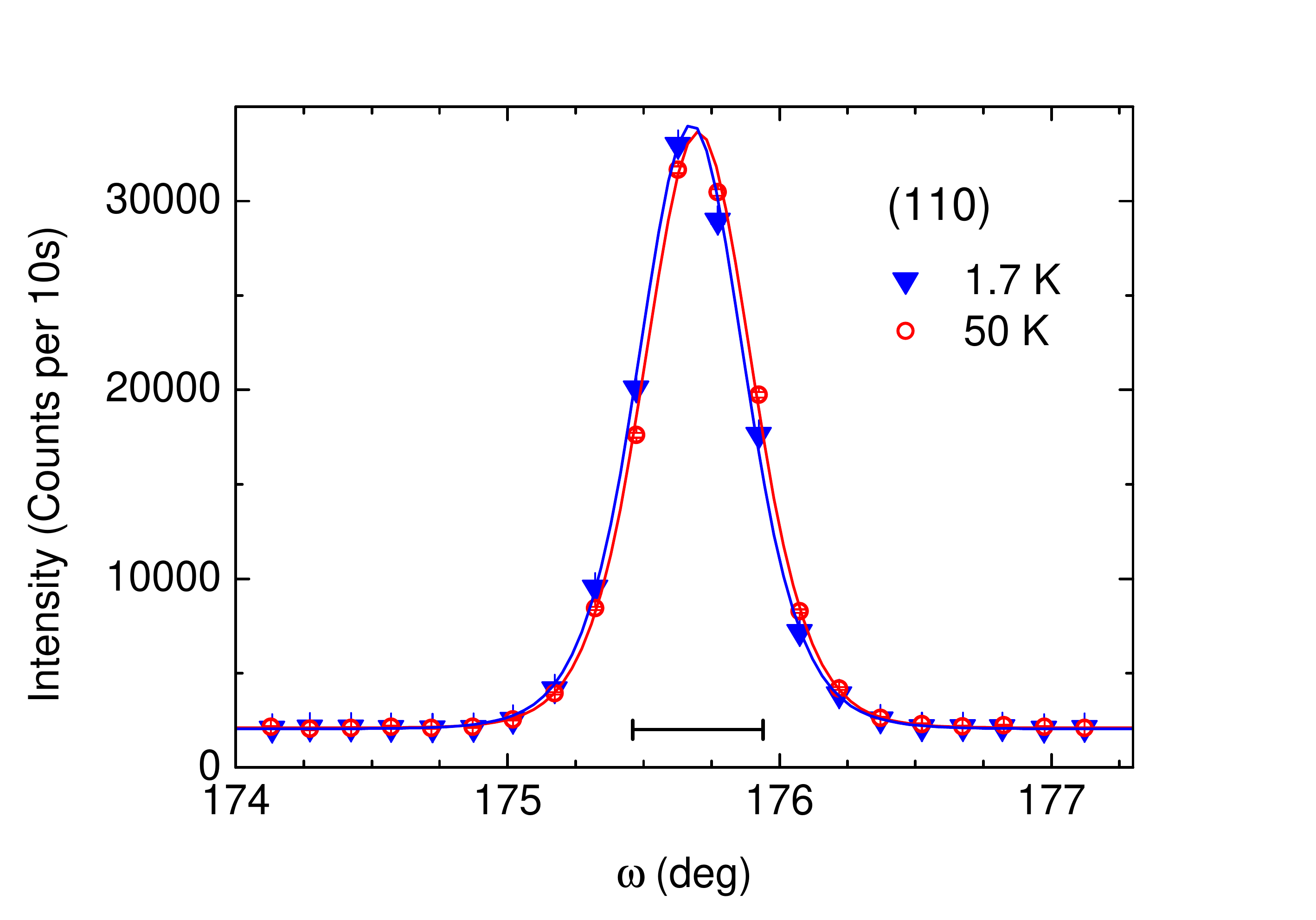}
\caption{(Color online) Rocking curves through the 110 nuclear reflection as recorded on E4 at 50 K and 1.7 K. The line segment at the bottom denotes the resolution of the E4 diffractometer at the scattering angle of the 110 reflection. Full lines through the experimental points are the best fits to a Gaussian profile. Note that the profile of the reflection is clearly resolution limited documenting a high quality of the single crystal.} \label{fig1}
\label{fig1}
\end{figure}

\section{Experimental}

Single crystalline \urxrs~has been prepared by a modified Czochralski method from a stoichiometric melt in high purity Argon (6N) atmosphere. No weight loss was detected, allowing us to denote the nominal Rh concentration as 8 \%. The maximal diameter was about 7 $mm$ with the length of about 50 $mm$. No further heat-treatment was applied after the crystal growth. The Laue X-ray backscatter images revealed very sharp reflections, along the $c$-axis, typical of four-fold symmetry. This method has also been used also to orient the crystal. We cut the oriented crystal using spark erosion in three pieces of different dimension that were used either for bulk measurements or neutron diffraction. Two of them are shown in the left inset of in Fig. \ref{fig2}.

Magnetization curves $M$($T$) with the magnetic field applied along principal directions were measured in the temperature range between 1.8 K and 350~K using the Quantum Design 14 T Physical Properties Measurements System (PPMS), which is part of the Laboratory for Magnetic Measurements at HZB. The static magnetic susceptibilities were calculated according $\chi$=$M$/$H$, where $H$ denotes the field strength from $M$($T$) dependencies.

Magnetic measurements were extended down to 400 mK using $^{3}$He fridge and a sensitive home-made ac susceptometer in a frequency range between 10 Hz and 10 KHz together with two magnet coil systems in order to eliminate possible artifacts originating from the detection system. 

Specific heat was measured on two single crystal pieces  (12.5 and 3.5 mg, respectively) in zero field between 1.8 K and 100 K in the same PPMS using the heat capacity option in fields up to 14 T. 

Neutron diffraction data were collected on the E1, E2, E4 and E5 instruments installed at the BER II 10 MW reactor of the Helmholtz-Zentrum Berlin. While the triple-axis instrument E1 and the normal-beam E4 diffractometer use the incident wavelength $\lambda$ = 2.41 \AA~produced by the PG (002) monochromator, on the flat cone diffractometer E2 we have used $\lambda$ = 1.21 \AA~produced by the Ge (311) monochromator. All the three instruments were used to characterize the quality of our single crystal and to follow and quantify the temperature evolution of the short-range signal between 1.7 K and 30 K. The E5 four-cycle diffractometer with Cu-monochromator selecting the neutron wavelength  $\lambda$ = 0.896 \AA~was used to collect a large data set to determine the structural details. 

The triple-axis instrument E1 is equipped with a single $^{3}$He-detector tube and PG analyzer leading good detection rates and a possibility to filter-out inelastic processes. We have used this instrument to separate a possible inelastic scattering contributing from the short-range signal.

The E2 flat-cone diffractometer is equipped with four two-dimensional position sensitive $^{3}$He-detectors (300 x 300 mm$^{2}$). Two wide-angle rocking curves with the four detectors shifted to fill gaps between them enable an  effective mapping and detection of all the diffracted signal not only within the scattering plane but also at significant distance above and below it.

The E4 diffractometer is equipped also with a two-dimensional position sensitive $^{3}$He-detector. Its size of 200 x 200 mm$^{2}$ makes it suitable to follow the selected signal as a function of an external parameter, in our case the temperature.

The E5 data were collected with a two-dimensional position sensitive $^{3}$He-detector, 90 x 90 mm$^{2}$ (32 x 32 pixels) at room temperature using a single crystal with the dimensions 4 x 4 x 4 mm$^{3}$ (see the smaller single crystal shown in the left inset of in Fig. \ref{fig2}). The crystal structure refinement was carried out with the program Xtal \cite{urrs1}. The nuclear scattering lengths b(Ru) = 7.21 fm, b(Rh) = 5.88 fm, and b(U) = 8.417 fm were used.\cite{urrs2} For the absorption correction (Gaussian integration) we used the absorption coefficient $\mu$ = 0.380 cm$^{-1}$. Secondary extinction has been corrected using the formalism of Zachariasen (type I). 

On E1, E2 and E4, we have used both, the large and smaller single crystals in the form of a truncated cone and a semi cube, respectively  (see the left inset in Fig. \ref{fig2}).

\section{Results}

\subsection{Crystal Structure}

A wide-angle rocking scans with the [100], [001] and [$\bar{1}$10] directions vertical collected using E4 diffractometer revealed that except for additional short-range magnetic correlations present at low temperatures indexable with \textbf{\textit{q$_ {III}$}} = ($\frac{1}{2}$ $\frac{1}{2}$ $\frac{1}{2}$), all observed Bragg reflections are compatible with the space group $I4/mmm$. Visible reflections are very sharp and resolution limited. Neither of the nuclear reflections show sizable thermal variation. As an example we show in Fig. \ref{fig1} rocking curves through the 110 nuclear reflection as recorded on E4 using the larger \urxrs~single crystal at 50 K and 1.7 K. The small shift in position is caused by the thermal expansion of the 
.

The 100 reflection, that is for the paramagnetic space group $I4/mmm$ forbidden and which can appear only as a consequence of an AF order (or crystal structure transformation) has not been observed at high temperature nor at 1.7 K even after collecting the data for several hours. The analysis based on the statistical error analysis \cite{urrs15} suggests that any dipole-like moment associated with SMAF order in our sample has to be smaller than $\approx$ 0.008 $\mu_{B}$.

\begin{figure}
\includegraphics[scale=0.32]{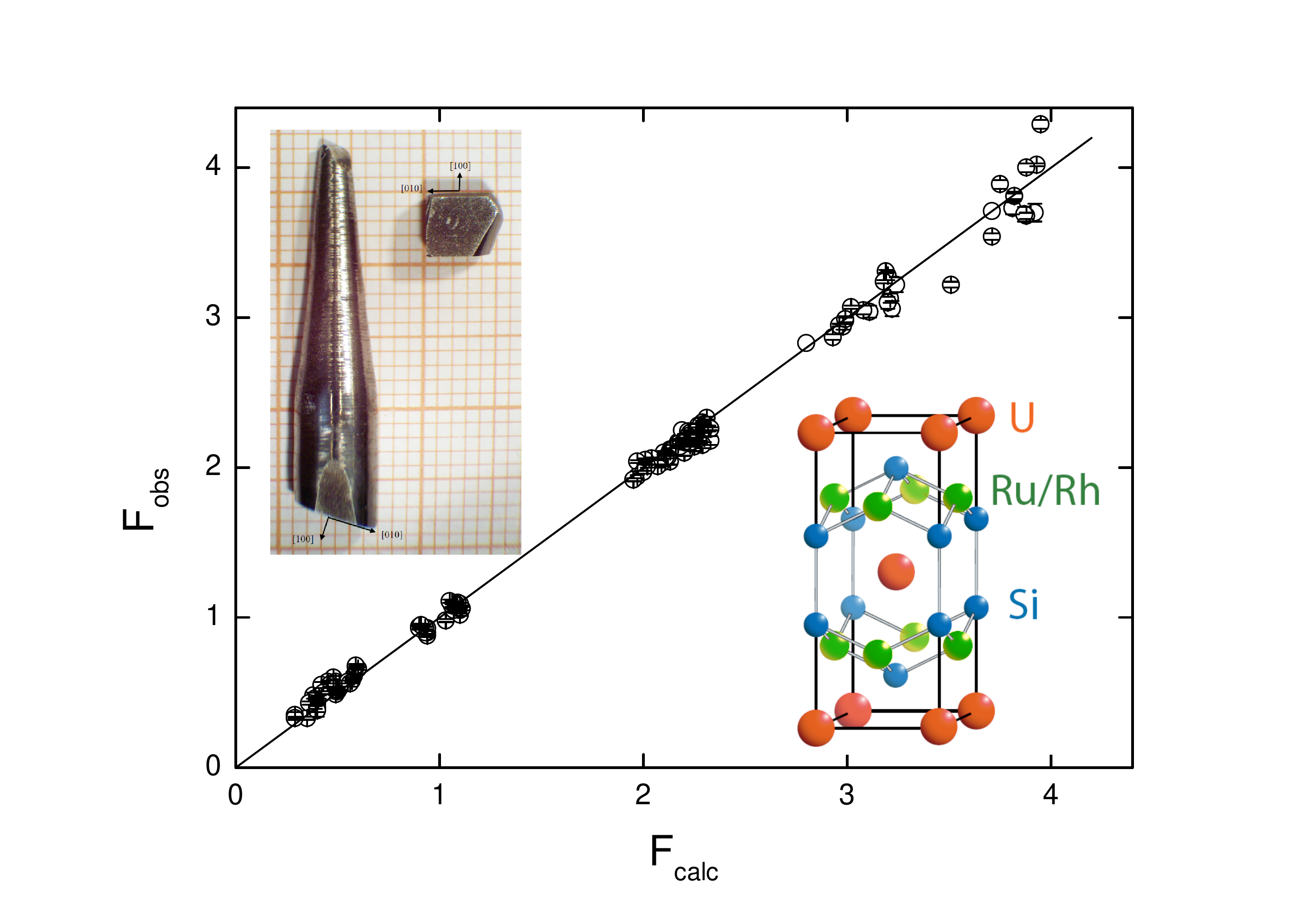}
\caption{(Color online) Plot of the observed versus calculated nuclear structure factors of \urxrs~after the extinction correction (E5 data taken at room temperature). A schematic representation of the \tcs~crystal structure adopted by \urrs~ is shown in the right inset. The atoms are shown as large (red), intermediate (green) and small (blue) circles stand for U, Ru/Rh and Si, respectively. The two single crystal used in the diffraction experiments are shown in the left inset. }
\label{fig2}
\end{figure}

The details of the crystal structure at room temperature were determined from E5 data set that contained 664 individual reflections (144  unique ones). It has been confirmed that \urxrs~crystallizes in the \tcs~type of structure with the tetragonal space group $I4/mmm$ (No. 139). In this space group are the U, Ru(Rh), and Si atoms at the Wyckoff positions 2$a$(0,0,0), 4$d$(0 ,$\frac{1}{2}$, $\frac{1}{4}$), and 4$e$(0,0,$z$), respectively. The $z$ parameter for Si atoms is the only free positional parameter. The refinement of the scale factor, the positional parameter $z$(Si), and the anisotropic thermal parameters of the different atoms resulted in the residuals $R_{F}$ = 0.030 defined as $R_{F}$ = $\sum$($\mid\mid$F$_{o}^{2}\mid$ - $\mid$F$_{c}^{2}\mid\mid$) / $\sum$($\mid$F$_{o}^{2}\mid$).  For the extinction parameter $g$, which is related to the mosaic distribution, we obtained the value $g$ = 472(24) rad$^{-1}$. The observed versus calculated structure factors are shown in Fig.~\ref{fig2}. As can be seen, a satisfactory agreement is obtained. The numerical results of the refinements are summarized in Table~\ref{tab:table1}. Using the lattice parameters $a$ = 4.0927(6) \AA~and $c$ = 9.5387(16) \AA~ the shortest U-U neighbor are found along the $a$-axis with separation equal to the $a$-axis lattice parameter. There are four of such neighbors. Further eight U next-nearest neighbors are found along the body diagonal at a distance of 5.5786(10) \AA. 
Both principal lattice parameters appear to be slightly smaller  than parameters of the pure \urs~\cite{urrs28,urrs29,urrs30}, what is surprising in view of the effect of Rh doping under which the $a$-axis parameter shortens and the $c$ parameter expands\cite{urrs12, urrs31}. The most sensitive parameter of Rh doping is thus a change in the $c$/$a$ ratio which should increase with increasing Rh content. Indeed, this ratio is for our sample by about 0.5 \% larger than for pure \urs. We therefore attribute the discrepancy in absolute values of lattice parameters to the uncertainty of the incident neutron wavelength. However, also the positional parameter of Si is slightly different from a value of 0.3710 listed in the literature for the pristine \urs~ determined at similar conditions. Here, however, no information on the Rh doping effect is reported so far. The calculated bond distance $d$(Si-Si) amounts to 2.412(1) \AA~ which is $\approx$  2 \% less than for the pure \urs. Let us note that we were unable to determine reliably the stoichiometry of our sample due to strong correlation among other free parameters, especially with extinction and thermal parameters. However, we could confirm that the starting 8 \% stoichiometry is compatible with our neutron crystallographic data.

\begin{table}
\caption{\label{tab:table1}Crystal structure parameters of a \urxrs~single crystal as 
determined from the E5  diffraction data at room temperature.  The thermal parameters $U_{ij}$ (given in 100 \AA$^{2}$) are in the form exp[-2$\pi^{2}$(U$_{11}h^{2}a^{*2}$ + …2U$_{13}hla^{*}c^{*}$)], where $hkl$ are indices of the relevant Bragg reflection and $a^{*}$ and $c^{*}$ are reciprocal lattice constants.  For symmetry reasons the values $U_{12}$, $U_{13}$, and $U_{23}$ of the atoms U, Ru(Rh) and Si are equal to zero in this structure. }

\begin{tabular}{lcccccc}
  \hline
  \hline
T = & 297 K  &  &  & & & \\
S.G. & $I$4/$mmm$  & & & \\
  Atom & Site  &~~~$x$~~~&~~~$y$~~~& $z$ & $U_{11}$ = $U_{22}$ & $U_{33}$ \\
\hline
  U &  2a & 0 & 0 & 0 & 0.29(3) & 0.47(4) \\
  Ru/Rh &  4d & 0 & $\frac{1}{2}$ & $\frac{1}{4}$ & 0.24(3) & 0.62(3) \\
  Si &  4e & 0 & 0 & 0.37325(10) & 0.37(3) & 0.52(5) \\
 \hline
 \hline
  \end{tabular}
\end{table}

\subsection{Magnetic bulk measurements}

\begin{figure}
\includegraphics[scale=0.3]{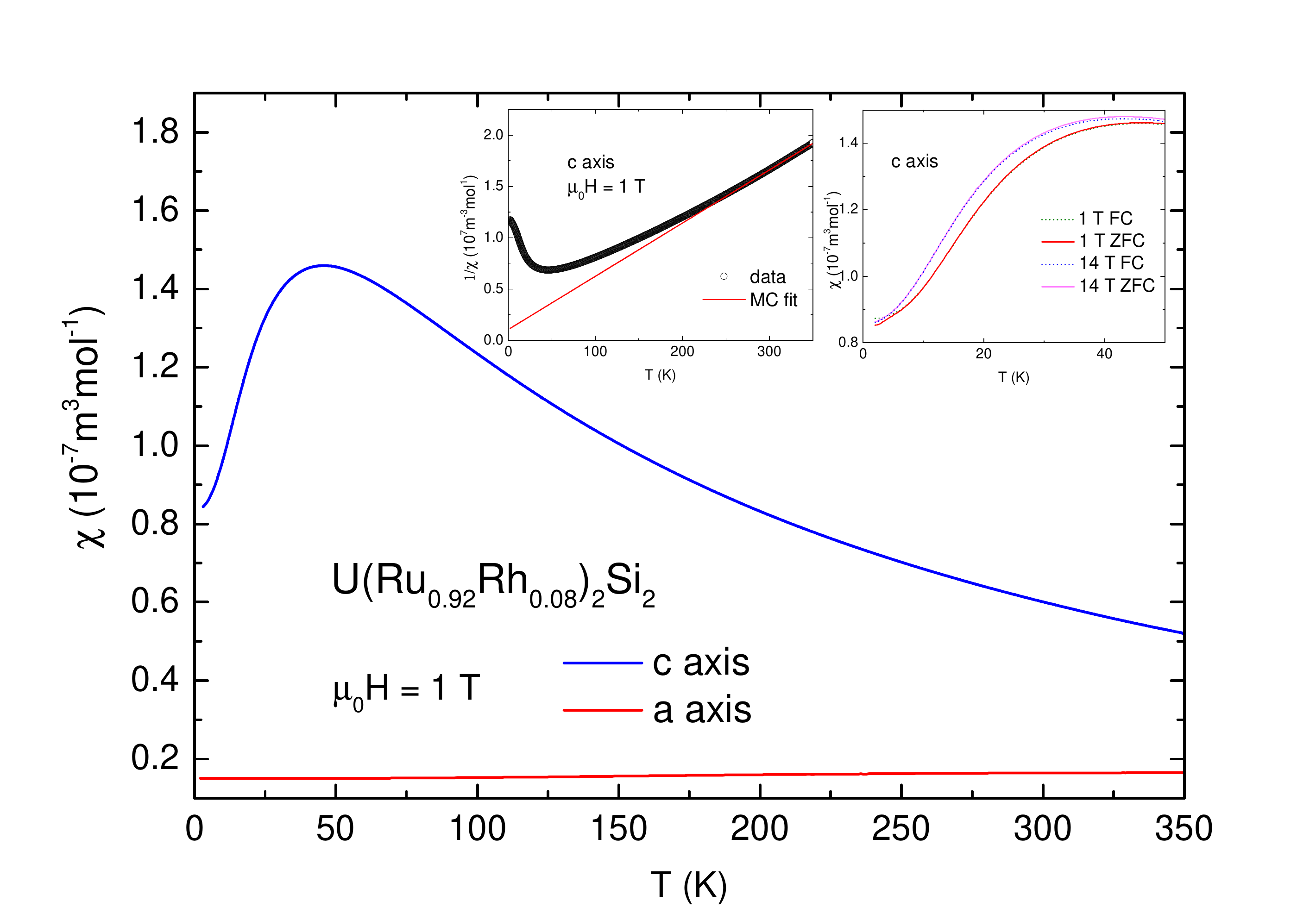}
\caption{(Color online) Temperature dependence of the static magnetic susceptibility $\chi$=$M$/$H$ of \urxrs~single crystal measured in 1 T applied along the principal axes. In the left inset we show the temperature dependence of the inverse magnetic susceptibility measured in 1 T applied along the tetragonal axis together with the best fit to a Curie-Weiss law to data above 300 K. In the right inset the low-temperature detail of the $c$-axis magnetic susceptibilities measured in 1 and 14 T with field applied along the tetragonal axis in zero-field cooled and field cooled regimes is shown. Note that there is no sharp shoulder around 17 K as in the case of the pristine \urs.}
\label{fig3}
\end{figure}

The temperature dependence of the \urxrs~static magnetic susceptibility $\chi$=$M$/$H$ measured in 1 T applied along the $a$ and $c$-axis is shown in Fig. \ref{fig3}. $\chi(T)$ is very anisotropic with the $a$-axis magnetic susceptibility $\chi_{a}$($T$) essentially temperature independent at a level of 0.16 10$^{-7}$ m$^{3}$mol$^{-1}$ in the whole range measured. For this direction no analysis according to a Curie-Weiss (CW) law $\chi_{c}$($T$) = $C$/($T$ - $\theta_{p}$), where $C$ denotes the Curie constant and $\theta_{p}$ the paramagnetic Curie temperature, is possible. For comparison, the $\chi_{c}$($T$) follows a CW law only in the high-temperature limit. Above 300 K, the best fit shown in the left inset of Fig. \ref{fig3} by the solid line through experimental $\chi_{c}$($T$) points yields $C$ = 1.9316 (5) 10$^{-5}$ and a negative $\theta_{p}$ = -20.6 (8) K. The effective magnetic moment per uranium atom of 3.51 (1) $\mu_{B}$ derived from the Curie constant is close to both, the effective moment found for the pure \urs~system along the $c$-axis and the effective moment expected for a localized U$^{3+}$ or  U$^{4+}$ configurations (3.58 and 3.62 $\mu_{B}$, respectively). 

\begin{figure}
\includegraphics[scale=0.32]{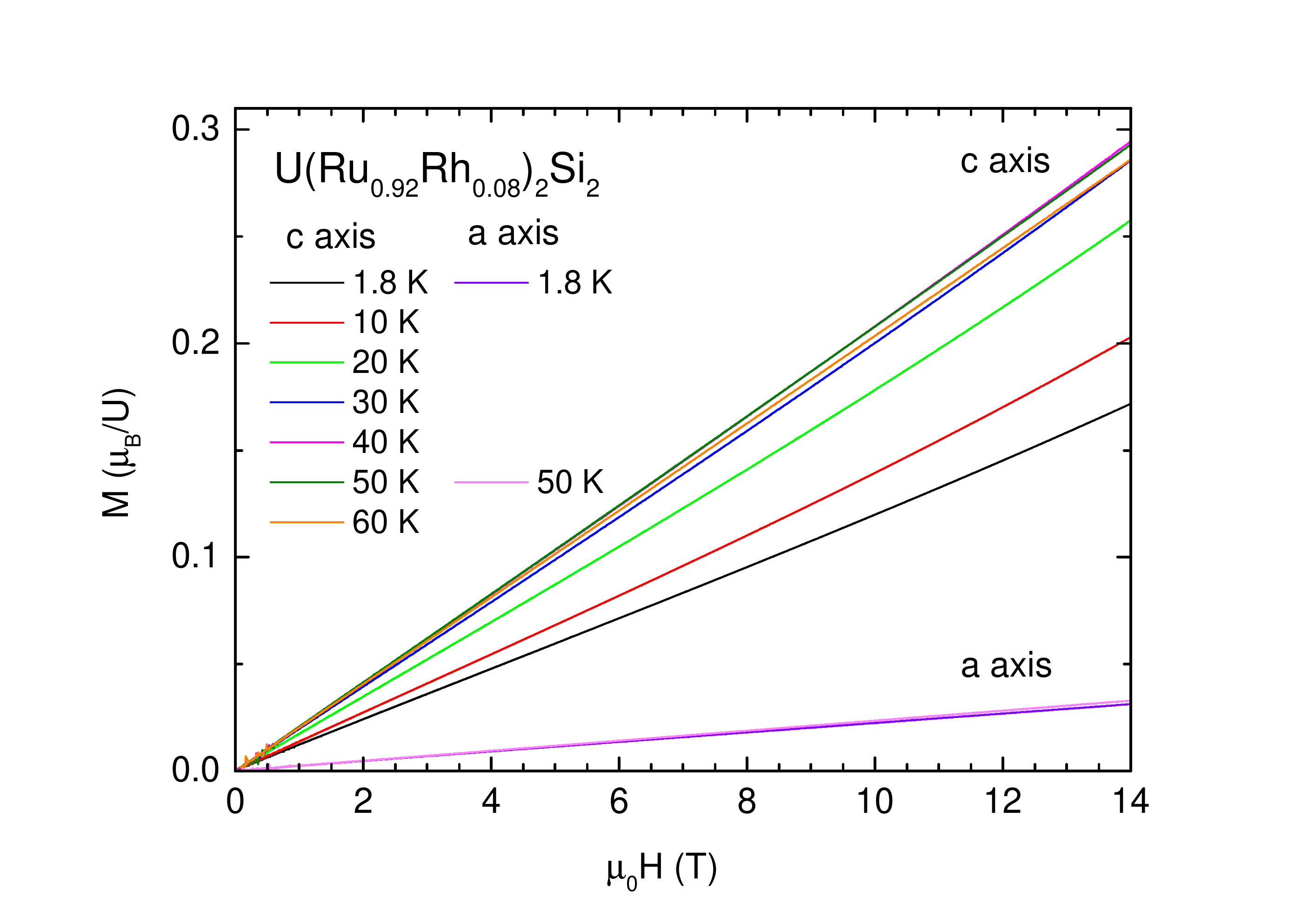}
\caption{(Color online) Field dependence of the \urxrs~single crystal magnetization measured at various temperatures between 1.8 K and 60 K with field applied along the $a$ and the $c$-axis. Note the nearly linear $M-H$ behavior throughout.}
\label{fig4}
\end{figure}

As the temperature is lowered, the $\chi_{c}$($T$) deviated progressively from a CW behavior, exhibits an maximum around 46 K and decreases strongly at lower temperatures. It has a weak smooth inflection point around 15 K and saturates eventually around a value of 0.85 10$^{-7}$ m$^{3}$mol$^{-1}$ in the low-temperature limit. This value is six times larger than the value found for $\chi_{a}$ at low temperatures and by $\approx$ 40 \% larger than $\chi_{c}$ at low temperatures observed for pristine \urs. 

The magnetic susceptibility along the $c$-axis, $\chi_{c}$($T$) is in the high-temperature limit independent of the applied field. At lower temperatures, however, differences are visible. First, the temperature at which the $\chi_{c}$($T$) exhibits maximum shifts slightly to lower values. At 14 T it is found around 43 K. Second, $\chi_{c}$($T$) shows slight field dependence. Its values measured at 14 T are generally lower above 60 K and larger below this temperature with respect to 1 T values. At lower fields, there is also a history dependence manifested by a tiny splitting of zero filed cooled (ZFC) and field cooled (FC) $\chi_{c}$($T$) curves measured at 1 T that merge again around 10 K. With increasing field the difference between the FC and ZFC curves vanishes. These observations are documented in the right inset of Fig. \ref{fig3}.

AC susceptibility measurements down to 400 mK performed in a wide frequency range did not revealed any anomalies that would suggest any HO, magnetic or superconducting phase transition. So, although the overall temperature dependence of both, $\chi_{a}$($T$) and $\chi_{c}$($T$) resembles very strongly the behavior of \urs, no phase transitions down to 400 mK can be detected in the case of \urxrs.

In Fig. \ref{fig4} we show the field dependence of the magnetization measured along the $c$-axis ($a$-axis) between 1.8 K and 60 K (50 K) in fields up to 14 T. Clearly, the response along the $c$-axis is much larger than along the $a$-axis. The response to the magnetic field along the $a$-axis is nearly temperature independent in the measured range, whereas for the $c$-axis it is nearly a factor of two stronger than at 1.8 K. These findings are in agreement with magnetic susceptibility results and establish a huge magnetocrystalline Ising-like anisotropy in \urxrs~ similar to \urs.  Absence of any anomalies up to 14 T along both directions suggest that \urrs~remains paramagnetic even at 14 T. However, we note that the magnetization does not increase strictly linearly with field. There is small but finite concave shape of magnetization curves up to 50 K (compare in Fig. \ref{fig4} for instance magnetization recorded at 40 K and 50 K with curve at 60 K). For the $a$-axis direction is the response linear at all temperatures.

\subsection{Electrical Resistivity}

\begin{figure}
\includegraphics[scale=0.30]{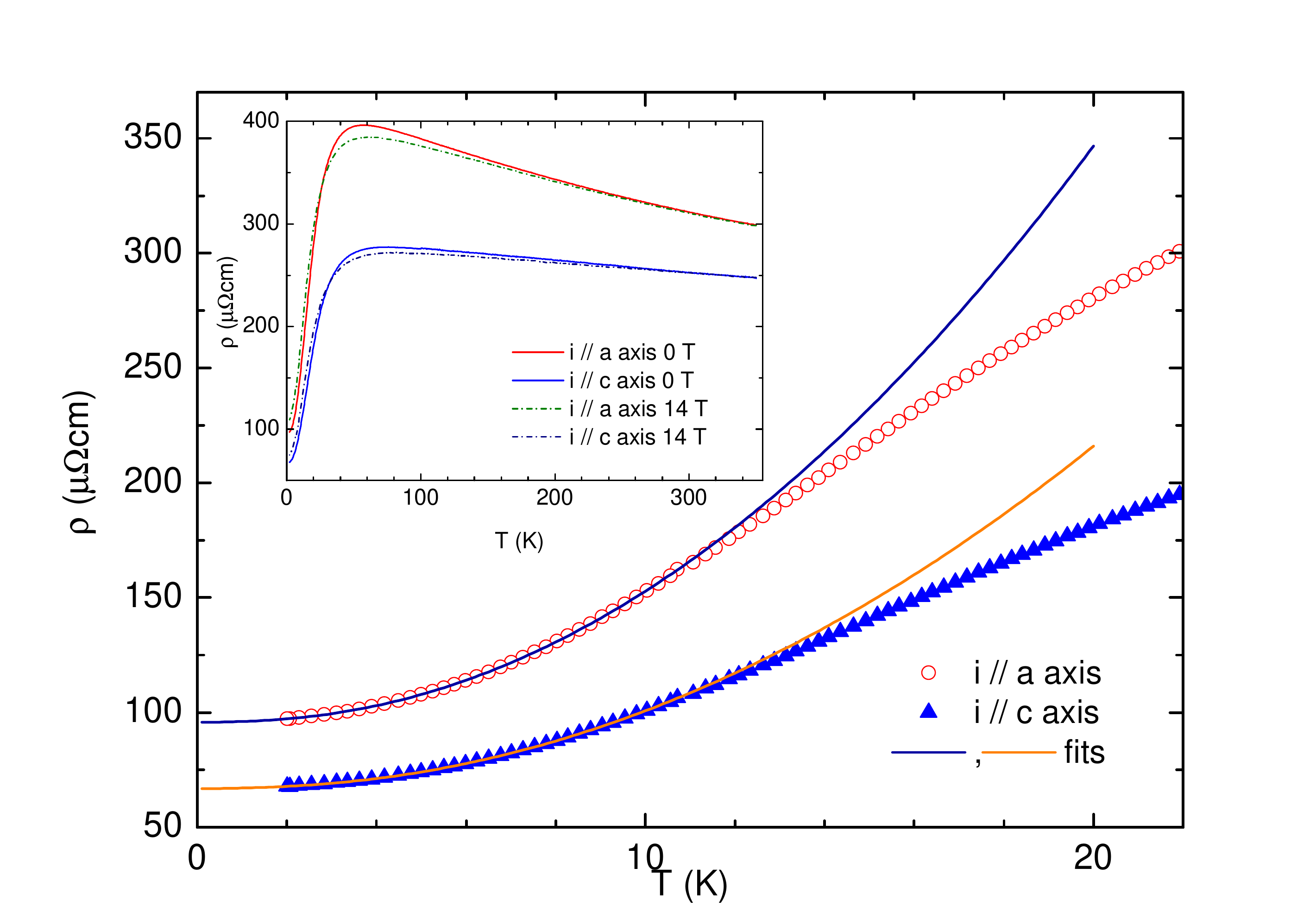}
\caption{(Color online) The zero-field low-temperature dependence of the \urxrs~electrical resistivity for current along the $a$ and the $c$-axis together with the best fit described in the main text. In the inset the electrical resistivity measured along the $a$-axis and the $c$-axis in the whole temperature range in zero field and 14 T applied along the  tetragonal axis is shown. }
\label{fig5}
\end{figure}

In the inset of Fig. \ref{fig5} we show the electrical resistivity for current along the $a$ and the $c$-axis measured in zero and 14 T external field applied along the tetragonal axis in the temperature range between 2 and 350 K. $\rho(T)$ is along both directions rather large at high temperatures and very anisotropic in the whole temperature range. At 350 K, the electrical resistivity measured along the $c$-axis, $\rho_{c}$, reaches $\approx$ 240 $\mu\Omega$cm$^{-1}$, while the $a$-axis value $\rho_{a}$ is about 40 \% larger. Both, $\rho_{c}$($T$) and $\rho_{a}$($T$) increase upon lowering the temperature indicating a formation of a heavy-fermion state, exhibit a maximum at slightly different temperatures and decrease strongly below 50 K to level-off in the low-temperature limit. No anomalies indicating phase transitions are visible. While $\rho_{a}$($T$) shows a maximum at 55 K, $\rho_{c}$($T$) peaks at higher temperature of about 75 K. The ratio $\rho_{a}$/$\rho_{c}$  increases upon lowering the temperature very slightly and attains a maximum around 55 K. The high-temperature anisotropy is smaller than the anisotropy found for \urs~\cite{urrs3} (where $\rho_{a}$ is about twice as large as $\rho_{c}$), however, it remains in \urxrs~finite also in the low-temperature limit, where the anisotropy for the pristine compound vanishes. The residual resistivity is large most probably due to the disorder caused by the Rh substitution.

The low-temperature parts with current along both directions, shown in the main panel of Fig. \ref{fig5}, cannot be described by perfect Fermi-liquid dependence of the form $\rho$($T$) = $\rho_{0}$ + $a T^{n}$ , $n$ = 2.0. The best fit to data between 2 and 10 K yields $n$ $\approx$ 2.2 for both directions. Even better agreement with data in the same temperature range is obtained for expression $\rho$($T$) = $\rho_{0}$ + $a T^{2}$ + $bT$(1 + 2$T$/$\Delta$)e$^{-\Delta/T}$. This formula has been introduced in order to account for the influence of an energy gap $\Delta$ in the dispersion relation of magnetic excitations caused by strong electron-magnon coupling \cite{urrs16} and was also used in order to describe the electrical resistivity behavior of pure \urs~\cite{urrs32}. The best fits to this formula between 2 K and 10 K are shown in Fig. \ref{fig5} by solid lines through the experimental points. The smooth $\rho$($T$) behavior eliminates any possibility of HO or SC phase transitions. For numerical results see Appendix.

\subsection{Magnetoresistance}

\begin{figure}
\includegraphics[scale=0.30]{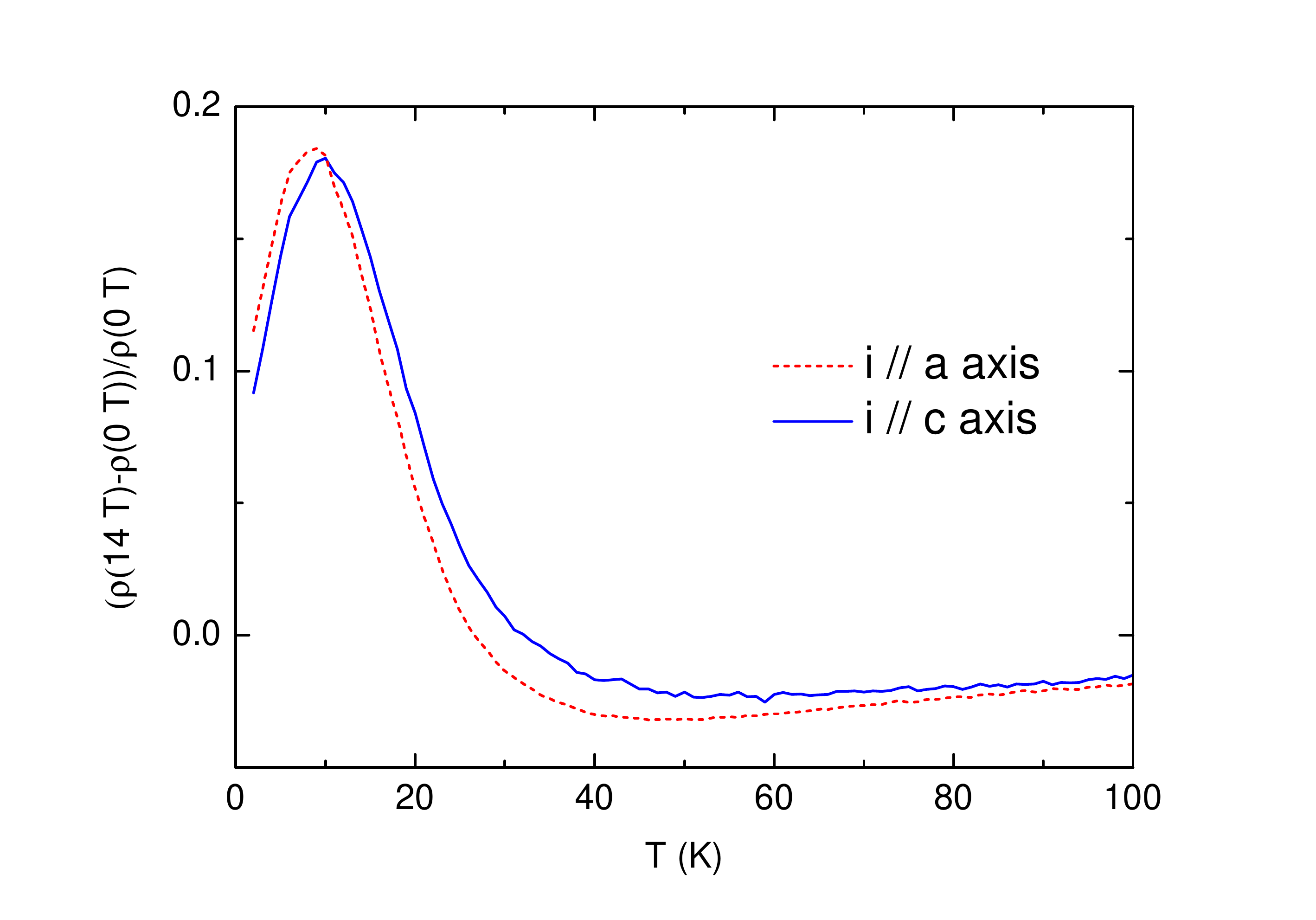}
\caption{(Color online) The temperature dependence of the net magnetoresistance between 14 T applied along the tetragonal axis and the zero-field state between 2 and 100 K. }
\label{fig6}
\end{figure}

The effects of the external field on the electrical resistivity of \urxrs~are very anisotropic. While the application of field along the hard magnetization $a$-axis leads to identical temperature dependencies, for the field applied along the $c$-axis we observe readily observable changes. This is displayed in the inset of Fig. \ref{fig5}. Small shifts of temperatures where maxima in $\rho$($T$) occur are detected. At 14 T the $\rho_{a}$($T$) and $\rho_{c}$($T$) maxima shift to about 60 and 73 K, respectively.  The field increases (decreases) the resistivity at temperatures below (above) 30 - 35 K along both directions. This leads to a magnetoresistance of opposite signs below and above this temperature. ($\rho_{0 T}$-$\rho_{14 T}$)/$\rho_{0 T}$ is shown in Fig. \ref{fig6}. For the 14 T it reaches 18 \% around 8-10 K for both directions. The detailed temperature dependence of $\rho$($T$) in magnetic fields is also modified. In 14 T (in contrast to zero field) it can be, however, described by $\rho$($T$) = $\rho_{0}$ + $a T^{2}$ dependence suggesting a full restoration of the heavy Fermi liquid behavior. For numerical results see Appendix. 

\begin{figure}
\includegraphics[scale=0.4]{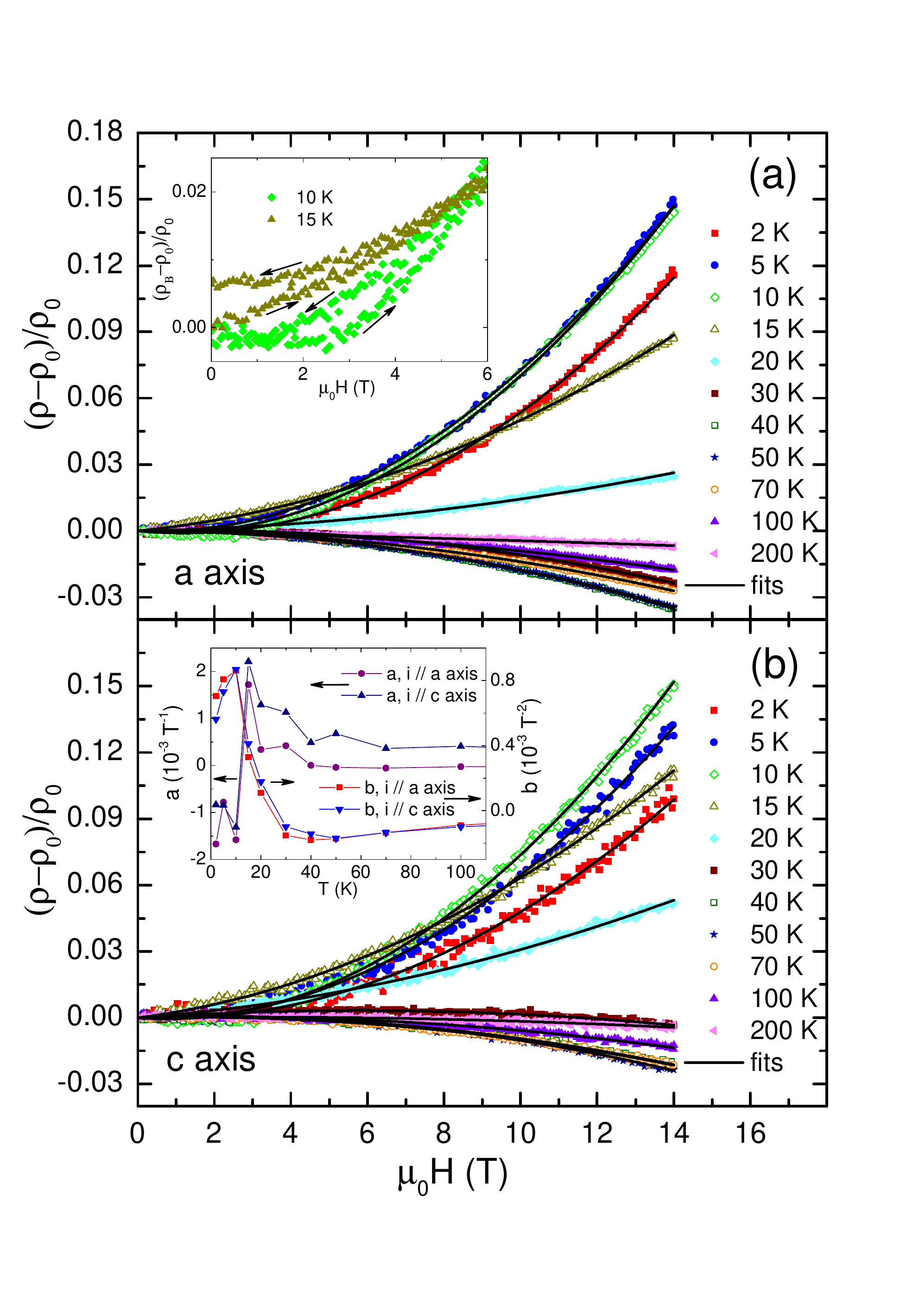}
\caption{(Color online) The field dependence of the \urxrs~single crystal measured with the current along the $a$-axis (a) and the $c$-axis (b) at various temperatures as a function of field applied along the $c$-axis. Full lines represent best fits to expression  ($\rho_{H}$ - $\rho_{0 T}$)/$\rho_{0 T}$ = $a H$ + $b H^{2}$. In the inset of panel (b) we show the temperature dependence of the fit parameters to this formula for both electrical current orientations. In the inset of panel (a) we document a hysteretic behavior for current along the $a$-axis that is found for both current orientations only at 10 and 15 K.}
\label{fig7}
\end{figure}

While there is nearly no field dependence along the $a$-axis, the electrical resistivity is strongly affected by a field applied along the $c$-axis. In Fig. \ref{fig7} we show the magnetoresistance of the \urxrs~single crystal measured with the current along the $a$ and $c$ axes at various temperatures as a function of field applied along the $c$-axis. Except for 10 and 15 K (shown in the inset  of Fig. \ref{fig6}a) there is no hysteresis between field sweeps up and down only increasing field curves are shown. Hysteresis is found for both current orientations only between 10 and 15 K. 

At low temperatures the electrical resistivity increases with the applied field $B$ = $\mu_{0}H$ and decreases at temperatures above 30 K (see also Fig. \ref{fig6}) for both current orientations. At low temperatures the field dependence cannot be described quadratically in the form ($\rho_{H}$ - $\rho_{0 T}$)/$\rho_{0 T}$ = $a H^{n}$ , $n$ = 2.0. This is in contrast to the pure \urs~ that is reported to exhibit the quadratic dependence \cite{urrs32}. Instead, our sample's magnetoresistance can be fit by an expression ($\rho_{H}$ - $\rho_{0 T}$)/$\rho_{0 T}$ = $a H$ + $b H^{2}$. The temperature dependencies of fit parameters from best fits are shown in the inset of of Fig. \ref{fig7}b. The ``a'' parameter (linear in field) is negative below 15 K and positive at all temperatures above this temperature. The ``b'' (quadratic term) is at first positive, increases with increasing temperature, peaks around 15 K and decreases above this temperature. It changes its sign, reaches large negative value around 40 K and decreases in absolute value with further increasing temperatures. Both parameters exhibit qualitatively the same behavior for the two current orientations and it can be seen that at temperatures above $\approx$~70-80 K the linear term diminishes and the ``c''-axis field dependence according to ($\rho_{H}$ - $\rho_{0 T}$)/$\rho_{0 T}$ = $a H^{n}$ with $n$ = 2.0 is restored.

\begin{figure}
\includegraphics[scale=0.33]{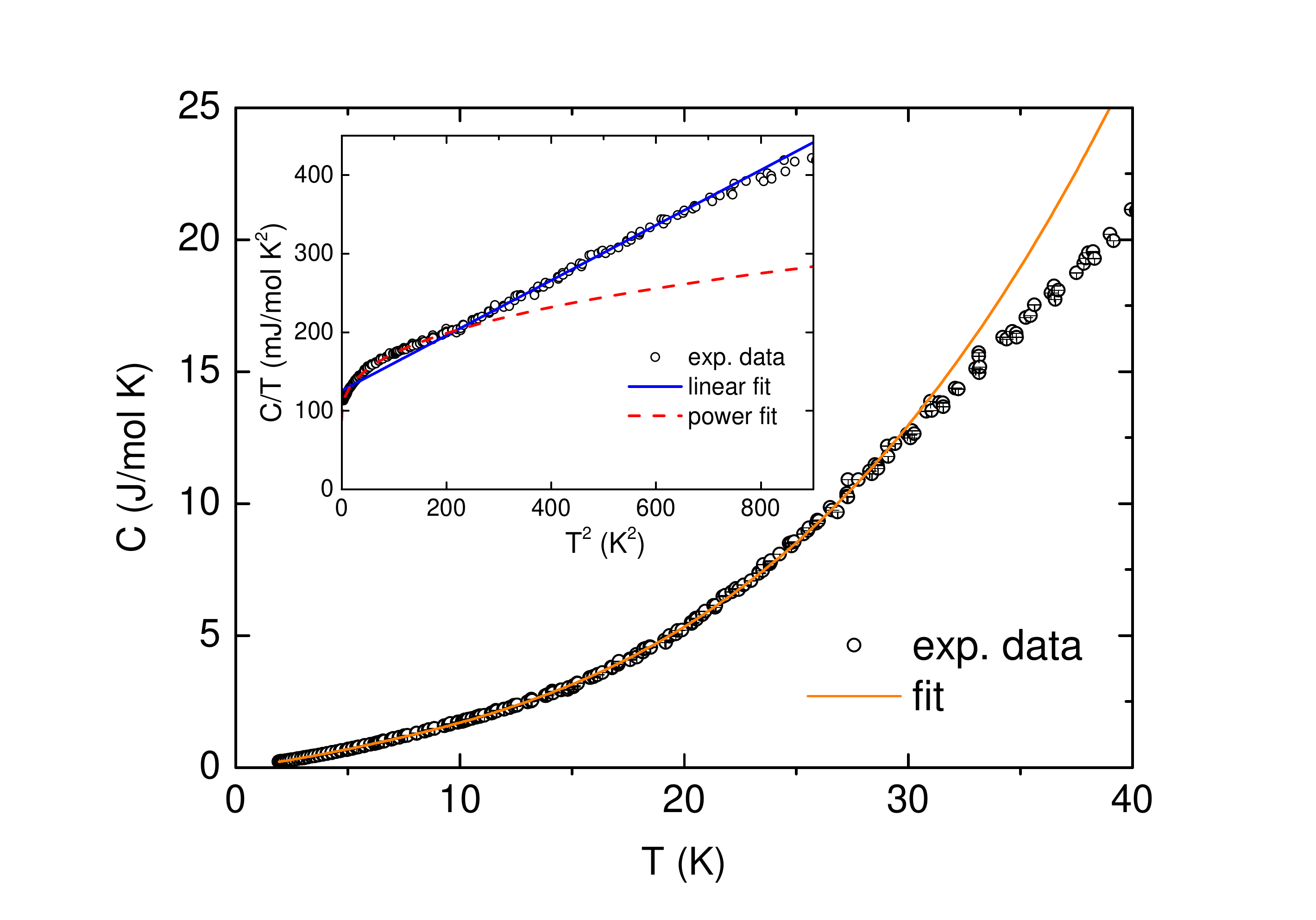}
\caption{(Color online) The temperature dependence of the \urxrs~specific heat together with the best fit containing an exponential term (see the main text). In the inset the low-temperature detail of the $C/T$ $vs.$ $T^{2}$ dependence is shown along with the best fits to a linear (solid, blue line) and power-law (dashed, red line) dependencies, respectively. }
\label{fig8}
\end{figure}

\subsection{Specific Heat}

The temperature dependence of the specific heat down to 1.8 K is shown in Fig. \ref{fig8}. No signature of any phase transition can be discerned  in the whole temperature range. This observation is in accord with the magnetic and electrical resistivity bulk measurements and document clearly the absence of the hidden order. The specific heat $C$ can be best fit between 1.8 K and 30 K to a formula $C$ = $\gamma T$ + $\beta T^{3}$ + $\delta$exp$^{-\Delta/T}$, where $\gamma$ denotes the electronic low-temperature specific heat coefficient, $\beta$ relates to the Debye temperature $\theta_{D}$ via expression $\theta_{D}^{3}$ = 12*$\pi^{4}R$/5$\beta$ and $\Delta$ denotes an energy gap in the putative dispersion relation of magnetic excitations \cite{urrs16}. The best-fit parameters to this formula yields  $\gamma$ = 111.9 (3.9) mJ /(mol K$^{2}$), $\theta_{D}$ = 181(1) K and $\Delta$ = 13.6 (7) K. The gap value is significantly lower than the value of 115 K found for the pure \urs~\cite{urrs3} and again suggests the absence of HO and SC in the present system. Let us note that the expression that neglects the gap term deviates from experimental data between 2 and 15 K. However, the deviation is small (less than 1 \%) and, at maximum, amounts to $\approx$ 15 mJ /(mol K). The difference between the ``non-magnetic background'' approximated by the formula without the gap term and measured specific heat can be interpreted as a contribution due to fluctuating magnetic moments. Magnetic entropy obtained by integration of this difference divided by temperature up to 15 K is tiny and amounts at most to 4*10$^{-4}Rln$2.

In the inset of Fig. \ref{fig8} we show the low-temperature detail of the $C/T$ $vs.$ $T^{2}$ dependence. As can be seen, experimental data show progressively downward curvature at the lowest temperatures. This is also a reason why it is not possible to deduce the low-temperature specific heat coefficient using the usually used expression $C/T$ = $\gamma$ + $\beta T^{2}$. However, above $\approx$ 17 K the data follow reasonably such a dependence.  The best fit to data between 17 and 27 K yields $\gamma$ = 125.6(1.6) mJ /(mol K$^{2}$) and $\theta_{D}$ = 177(2) K. This fit is shown in the inset of Fig. \ref{fig8} by the full (blue) line. The curvature of the data below 15 K can be accounted for by a power law in the form $C/T$ = $\gamma$ + $\beta T^{n}$. The best to this expression between the lowest temperature and 14.5 K yields a lower value of $\gamma$ = 92.2(1.7) mJ /(mol K$^{2}$) and n = 0.40(2) K.

\begin{figure}
\includegraphics[scale=0.33]{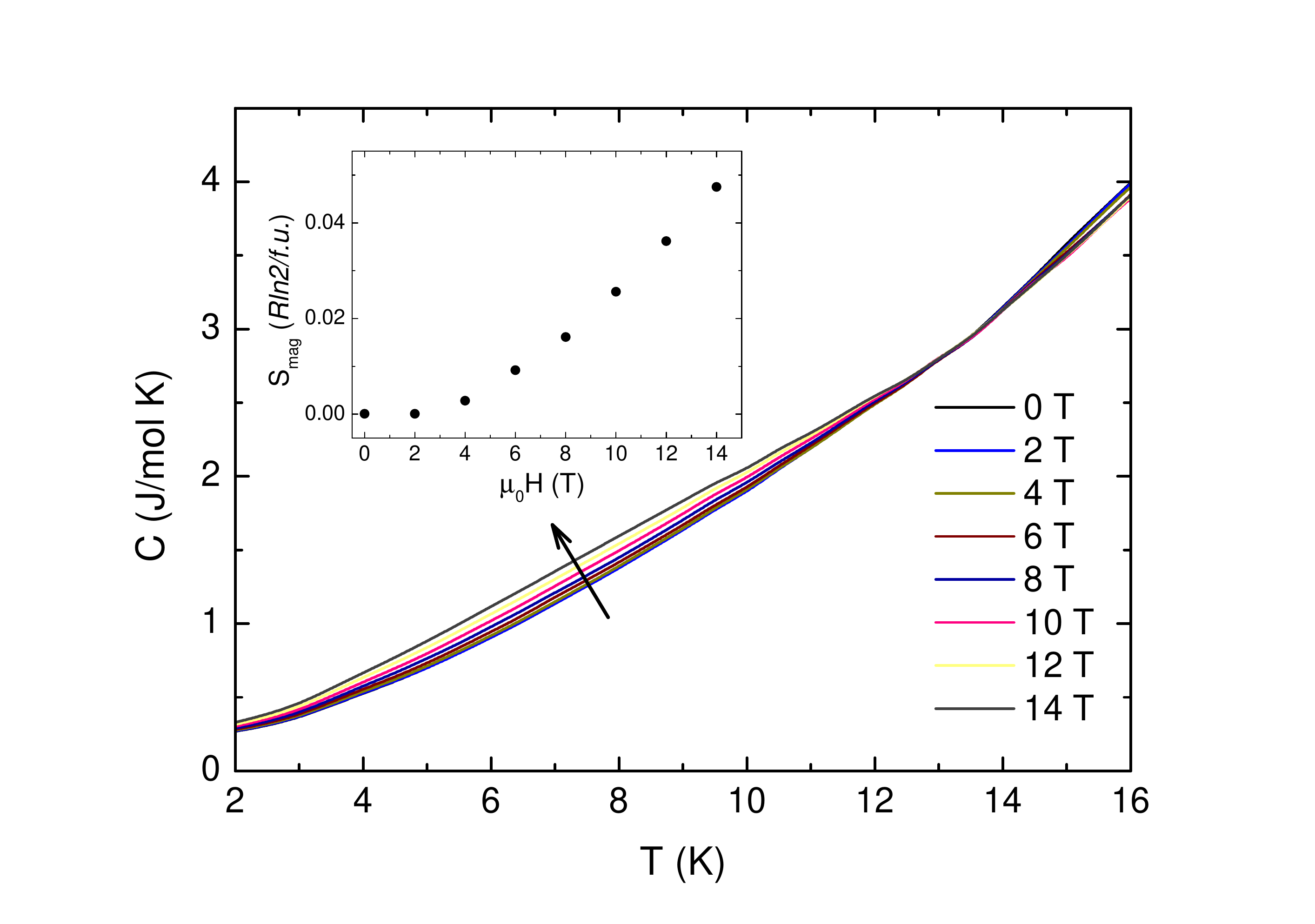}
\caption{(Color online) The temperature dependences of the \urxrs~specific heat measured at different magnetic fields applied along the tetragonal axis. In the inset the magnetic entropy calculated by integration of data shown in the main panel (from which an estimate of a non-magnetic analog contribution has been subtracted) divided by the temperature up to 14 K is shown.}
\label{fig9}
\end{figure}

In Fig. \ref{fig9} we show the low-T detail of the field dependence of the specific heat as a function of temperature measured at various magnetic fields applied along the $c$-axis. Such appears below 15 K $C(T, H)$ and increases with increasing field in a non-linear fashion. At constant temperature the field dependence can be described by a power law according to $C(T) \approx C_{0T}$ + $bH^{n}$ with $n$ dependent on the field. $n$ is close to 1.90 at 2 K and decreases slightly with increasing temperature. At 7 K it amounts to 1.7. Above this temperature the parameter $n$ decreases faster and approaches zero around 14 K. At higher temperatures the specific heat seems to decrease marginally with increasing field. Apparently, the magnetic entropy $S_{mag}$ calculated by integration of the specific heat from which the ``nonmagnetic'' background has been subtracted and divided by temperature between the lowest temperature and 14 K at particular field increases with increasing field (see the inset of Fig. \ref{fig9}. The dependence is again non-linear. At low fields $S_{mag}$ quadratically increase as a function of the applied field, above $\approx$ 4 T the dependence changes to a linear one. At 14 T the magnetic entropy amounts to 0.047 $Rln2$ per formula unit.

\begin{figure}
\includegraphics*[scale=0.23]{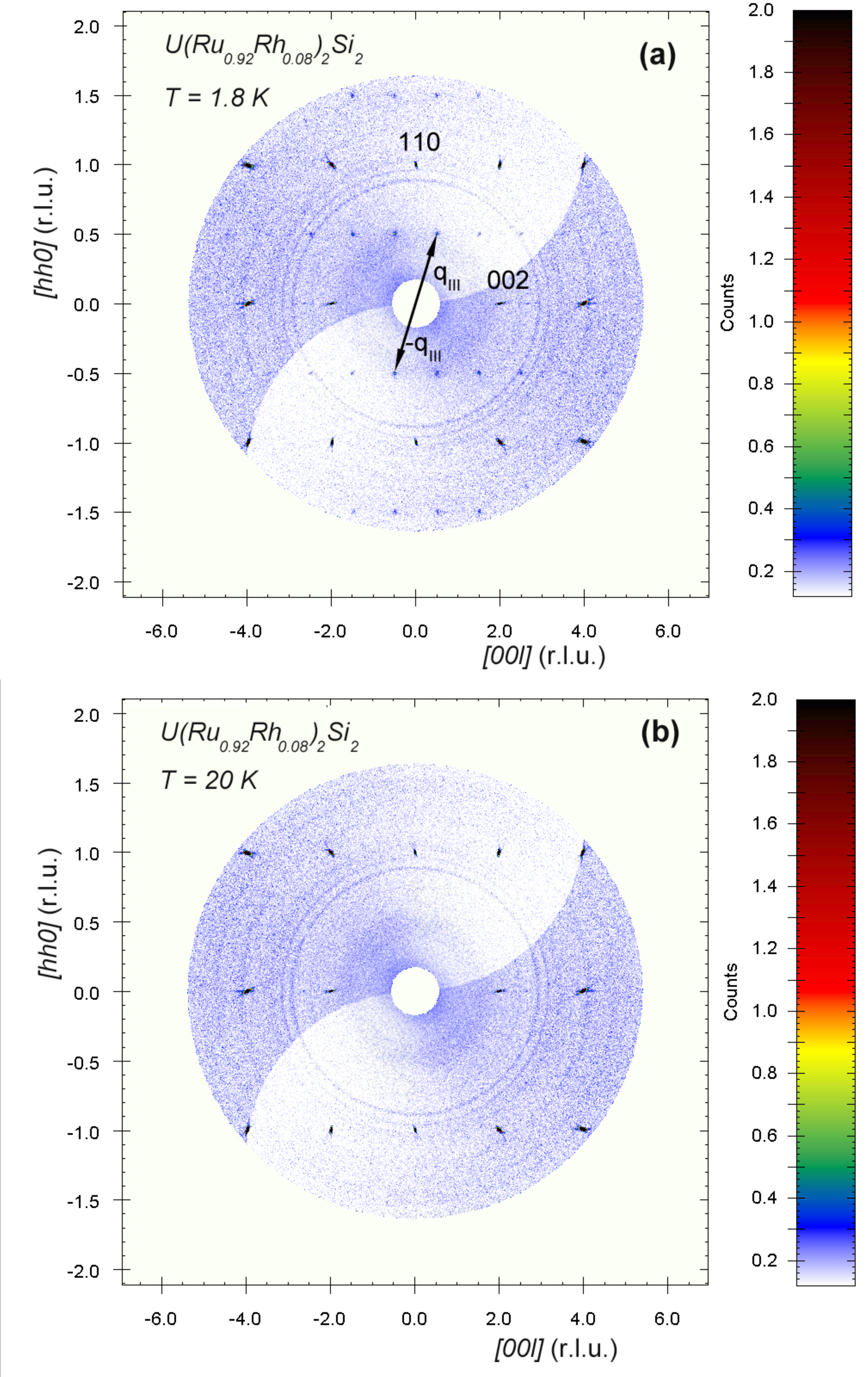}
\caption{(Color online) Diffraction patterns of the \urxrs~single crystal measured at 1.8 K (a) and at 20 K (b) in the ($hhl$) orientation using E2 diffractometer transformed into the reciprocal space. Note that this instrument senses a significant portion of the reciprocal space also above and below the scattering plane. Sharp strong nuclear Bragg reflections of the crystal main phase are observed along with broader features indexable with \textbf{\textit{q$_ {III}$}} = ($\frac{1}{2}$ $\frac{1}{2}$ $\frac{1}{2}$). \textbf{\textit{q$_ {III}$}} and -\textbf{\textit{q$_ {III}$}} are shown in (a) by arrows.} \label{fig10}
\end{figure}

\begin{figure}
\includegraphics*[scale=0.42]{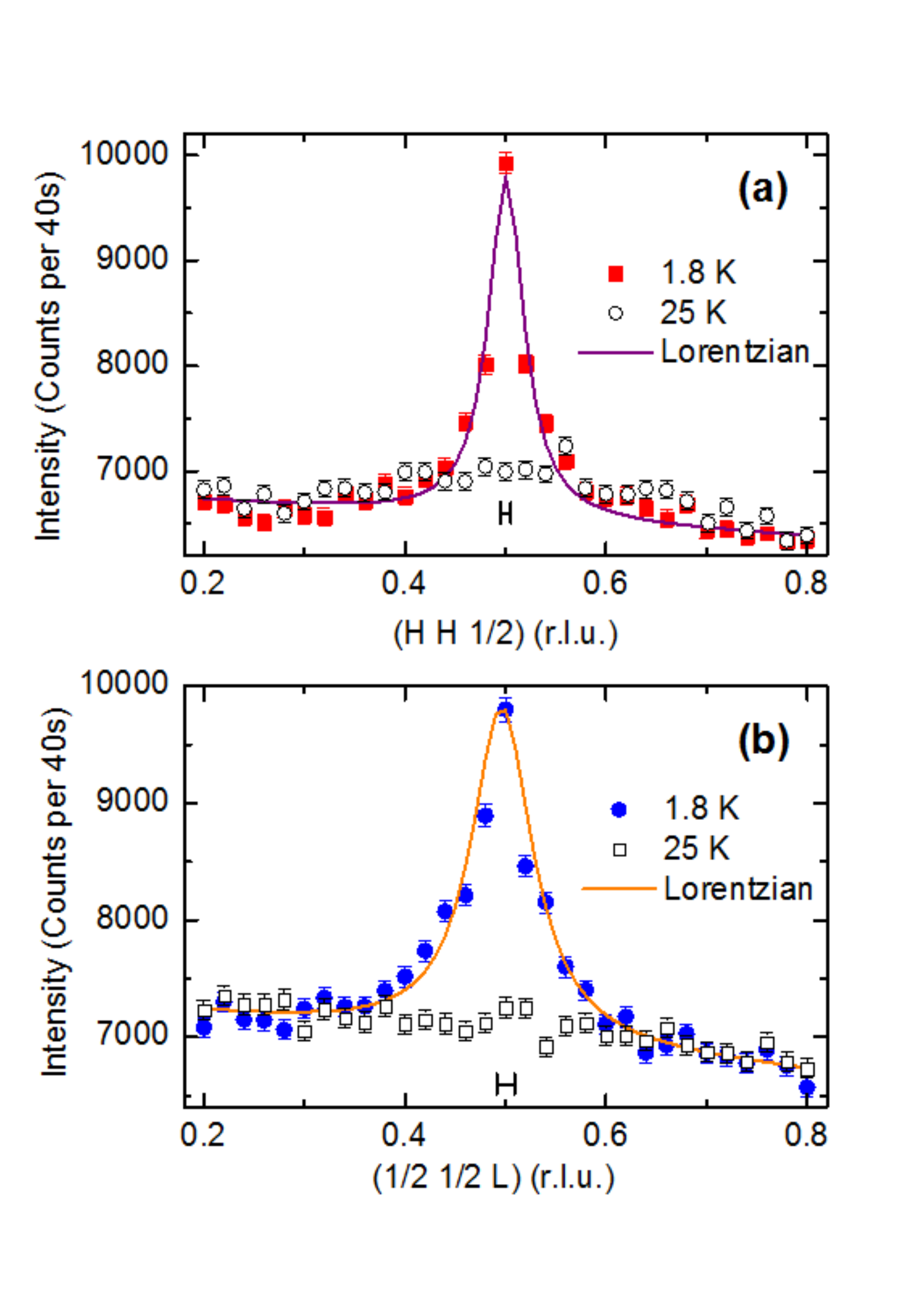}
\caption{(Color online) Diffraction pattern taken around the $h$ = $\frac{1}{2}$  $k$ = $\frac{1}{2}$  $l$ = $\frac{1}{2}$ position at 1.8 K and at 25 K as measured on the E4 diffractometer. (a) Reciprocal scan along the [$h$ $h$ 0] direction. The scan along the [0 0 $l$] direction is shown in panel (b). Full line through the experimental points are best fits to a Lorentzian profile. The short black line (``H'') at the bottom shows the instrument resolution at the current position.} \label{fig11}
\end{figure}

\subsection{Short Range Magnetic Order}

In Fig. \ref{fig10}(a) and (b) we show the reciprocal ($hhl$) plane of the \urxrs~single crystal as measured using neutron diffraction technique on the E2 instrument at 1.8 K and 20 K, respectively. As can be seen, at 2 K, apart from nuclear Bragg reflections, we also observe an additional, weaker and broader, signal at places that are indexable with \textbf{\textit{q$_ {III}$}} = ($\frac{1}{2}$ $\frac{1}{2}$ $\frac{1}{2}$). This signal has clearly Lorentzian profile and at low temperatures its full width at half maximum is three to four times larger than the width of the nuclear reflections that have Gaussian profiles. This fact is documented in Fig. \ref{fig11}(a) and (b) (please, compare with the 110 nuclear reflection shown in Fig.~\ref{fig1}), where we show representative reciprocal scans along [$h$ $h$ 0] and [0 0 $l$] directions through the ($\frac{1}{2}$ $\frac{1}{2}$ $\frac{1}{2}$) reciprocal space position as measured on the E4 diffractometer. Both scans can be fit to a Lorentzian profiles with the full width at a half maximum being about four times larger than the resolution. The correlation lengths are found to be $\approx$ 100 \AA~perpendicular to the $c$-axis and $\approx$ 200 \AA~along the tetragonal axis. This indicates that this magnetic signal is not due to a long-range magnetic order but short-range order (SRO). Such SRO has been previously observed in \urrs~single-crystalline sample with $x$ = 0.06 \cite{urrs12}. However, correlation lengths in our system are significantly larger. This agrees with the general tendency towards AFM order with increasing doping of Rh for Ru. With increasing Rh content beyond our 8 \% Rh crystal, the magnetically ordered phase appear and the transition temperature increase\cite{urrs12,urrs18}. In our system, however, no phase transition down to 0.4 K has been detected.

\begin{figure}
\includegraphics*[scale=0.32]{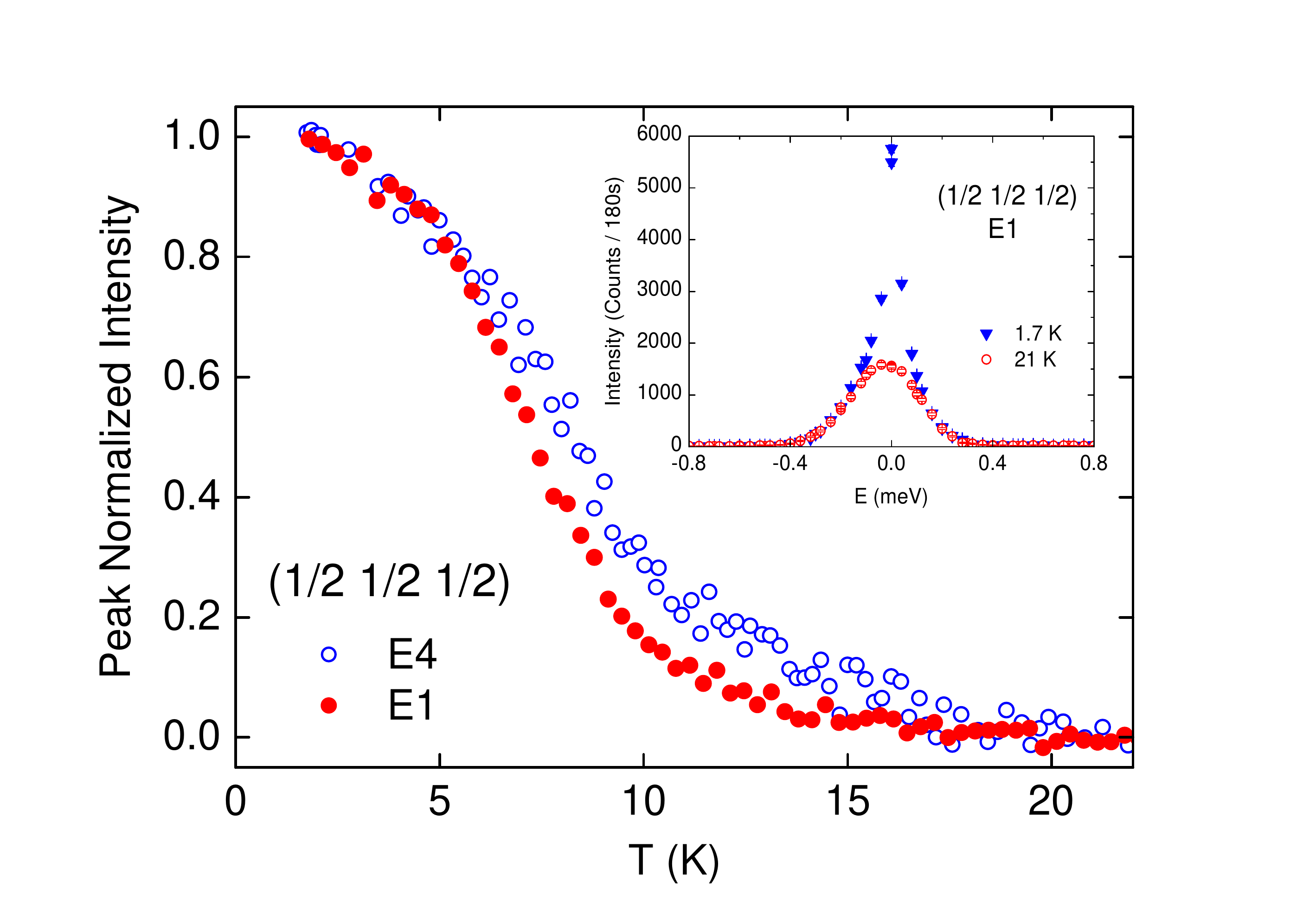}
\caption{(Color online) Temperature dependence of the diffraction intensity taken at the $h$ = $\frac{1}{2}$  $k$ = $\frac{1}{2}$  $l$ = $\frac{1}{2}$ reciprocal space position between 1.7 K and 21 K. In the inset we show an energy scan through this position using the tripple-axis spectrometer E1 at 1.7 K and 21 K. } \label{fig12}
\end{figure}

As the temperature is increased, the SRO intensity decreases. No traces of SRO could be observed neither in the E2 nor in the E1 or E4 data taken above 20 - 25 K. The temperature dependence of the peak intensity at the ($\frac{1}{2}$ $\frac{1}{2}$ $\frac{1}{2}$) position as measured on E1 and E4 is shown in Fig.~\ref{fig12}. As can be seen, it decreases with increasing temperature and exhibit an inflection point around 7-8 K with a long tail at higher temperatures. The background level is reached at lower temperatures on E1 as compared to E4 data. Nuclear Bragg reflections are shown in the same temperature range without a broadening or temperature dependencies. As the E4 instrument detects all the scattered neutrons irrespective of their energy, this observation suggests that some portion of detected neutrons at this reciprocal space position undergo inelastic scattering process. This is documented in the inset of Fig. \ref{fig12} which show the constant- \textbf{\textit{Q}} energy scan at the ($\frac{1}{2}$ $\frac{1}{2}$ $\frac{1}{2}$) position at 1.7 K and at 21 K recorded using the E1 spectrometer. It is apparent that the signal seen at the lower temperature is present within the whole energy window of the E1 spectrometer. Nuclear Bragg reflections, on the other hand, are much narrower in the energy and fit into this window entirely. This suggests that the magnetic signal is quasielastic in nature but extends most probably also outside the E1 energy window. This intensity, not detected at E1, is however collected on E4, thus causing the difference in the temperature dependences seen using the two instruments.

\section{Discussion and Conclusions}

Based upon our detailed and systematic bulk measurements of single crystal \urxrs~we have verified its heavy Fermi liquid behavior over the entire temperature and field ranges. However and most important, our 8 \% Rh substituted crystal does not exhibit any sort of phase transition or long-range order down to 0.4 K. Instead, according to our neutron scattering experiments, it displays short-range order of the \textbf{\textit{q$_ {III}$}} = ($\frac{1}{2}$ $\frac{1}{2}$ $\frac{1}{2}$) wave vector with a correlation length of ≈ 100 \AA (in plane) and ≈ 200 \AA (along the $c$-axis). The \textbf{\textit{q$_ {III}$}} propagation vector is distinctly different from the \textbf{\textit{q$_ {I}$}} = (1 0 0) characterizing the parasitic SMAF \cite{urrs20,urrs34,urrs35} (Fe or Os doping or pressure induced LMAF) or q$_ {II}$ = (0.6 0 0) propagation vector of the very high field spin density wave (SDW) observed in pure \urs~\cite{urrs33}. The long-range character of the \textbf{\textit{q$_ {III}$}} AF order appears at slightly higher Rh concentrations \cite{urrs12,urrs17}, not reached in our 8 \% Rh-doped crystal. Thus we have destroyed the HO and superconductivity with Rh substitution allowing a pure heavy Fermi liquid to remain. This then begs the question: How does this rather dilute Rh substitution (one extra 4$d$ electron) affect the band structure and Fermi surface?

\urxrs~is similar to the heavy fermion behavior of \urs~ in its field dependences at least up to our 14 T limit. Strong Ising- like anisotropy is formed along the $c$-axis with little or no field effects in the plane. Since there are no orderings in our crystal, the applied field only quantitatively induces small changes in the magnetization, specific heat and resistivity. Our maximum field of 14 T is insufficient to create the new magnetic state (\textbf{\textit{q$_ {II}$}}) that we expect to appear above 22 T. Here such high field magnetization and neutron experiments are planned. 

The interesting observation of  q$_ {III}$  short-range order that is absent in undoped \urs~ (where parasitic SRO with \textbf{\textit{q$_ {I}$}} is found) must be further studied. Our preliminary indications of magnetic fluctuations in the inelastic neutron scattering point to a precursor of the higher Rh concentration magnetic order found in Ref. \cite{urrs12,urrs17}. Finally we note the difference between Rh-tuning (leading to magnetic order with \textbf{\textit{q$_ {III}$}}), pressure tuning (magnetic order with \textbf{\textit{q$_ {I}$}}) and field tuning (characterized by \textbf{\textit{q$_ {II}$}}) of the HO behavior of \urs~towards LMAF: An intriguing problem for theoretical consideration.

\section{Appendix}

As the temperature dependence of the electrical resistivity deviates from the quadratic dependence, we have utilized formula~\cite{urrs32} $\rho$($T$) = $\rho_{0}$ + $a T^{2}$ + $bT$(1 + 2$T$/$\Delta$)e$^{-\Delta/T}$ that was used previously for the pristine \urs~\cite{urrs32}. These fits provide an excellent description of the temperature dependences both, along the $a$ and $c$ axis. The best fits between 2 and 10 K yield $\rho_{0,a}(0~T)$ = 95.7(2) $\mu\Omega$cm, $a_{a}(0~T)$ = 0.35(5) $\mu\Omega$cmK$^{-2}$ and $\Delta_{a}(0~T)$ = 7.7(1.3) K for the $a$-axis direction and $\rho_{0,c}(0~T)$ = 66.8(1) $\mu\Omega$cm, $a_{c}(0~T)$ = 0.25(2) $\mu\Omega$cmK$^{-2}$ and $\Delta_{c}(0~T)$ = 10.7(1.8) K for the $c$-axis direction, respectively. They are shown in Fig. \ref{fig5} by solid lines through the experimental points. Values of $\Delta$ parameters are for our sample significantly lower than for pristine \urs~\cite{urrs32}.

The field dependence of the electrical resistivity in 14 T applied along the $c$-axis is of the $\rho$($T$) = $\rho_{0}$ + $a T^{2}$ type. The best fit to this dependence between 2 and 10 K yields $\rho_{0,a}(14~T)$ = 106.4(3) $\mu\Omega$cm and $a_{a}(14~T)$ = 0.713(5) $\mu\Omega$cmK$^{-2}$ for the $a$-axis direction and $\rho_{0,c}(14~T)$ = 72.9(1) $\mu\Omega$cm and $a_{c}(14~T)$ = 0.442(3) $\mu\Omega$cmK$^{-2}$ for the $c$-axis direction, respectively.

\acknowledgments We acknowledge A. de Visser from University of Amsterdam for the general support of this project.


\begin{thebibliography}{4}

\bibitem{urrs3} T. T. M. Palstra, A. A. Menovsky, J. van den Berg, A. J. Dirkmaat, P. H. Kes, G. J. Nieuwenhuys,  J. A. Mydosh, $\it{Phys.~Rev.~Lett. }~ \textbf{55}$, 2727 (1985).

\bibitem{urrs14} K. Hasselbach, P. Lejay and J. Flouquet $\it{Phys.~Lett. A}~ \textbf{156}$, 313 (1991).

\bibitem{urrs4} J. A. Mydosh and P. M. Oppeneer, $\it{Rev.~Mod.~Phys.}~ \textbf{83}$, 1301 (2011).

\bibitem{urrs5} J. A. Mydosh and P. M. Oppeneer, $\it{Philos.~Mag.}~ \textbf{94}$, 3642 (2014).

\bibitem{urrs8} C. Broholm, J.~K. Kjems, W.~J.~L. Buyers, P. Matthews, T.~T.~M. Palstra, A.~A. Menovsky and J.~A. Mydosh
$\it{Phys. Rev. Lett.}~ \textbf{58}$, 1467 (1987).

\bibitem{urrs9} H. Amitsuka, M. Sato, N. Metoki, M. Yokoyama, K. Kuwahara, T. Sakakibara, H. Morimoto, S. Kawarazaki, Y. Miyako, and J. A. Mydosh $\it{Phys.~Rev.~Lett. }~ \textbf{83}$, 5114 (1999).

\bibitem{urrs10} K. H. Kim, N. Harrison, M. Jaime, G. S. Boebinger, and J. A. Mydosh $\it{Phys.~Rev.~Lett. }~ \textbf{91}$, 269902(E) (2003).

\bibitem{urrs11} M. Jaime, K. H. Kim, G. Jorge, S. McCall, and J. A. Mydosh $\it{Phys.~Rev.~Lett. }~ \textbf{89}$, 287201 (2002).

\bibitem{urrs26} Y. S. Oh, K. H. Kim, P. A. Sharma, N. Harrison, H. Amitsuka, and J. A. Mydosh $\it{Phys.~Rev.~Lett. }~ \textbf{98}$, , 016401 (2007).

\bibitem{urrs33} W. Knafo et al., private communication and to be published in $\it{Nat.Comm.}$~(2016).	

\bibitem{urrs12} P. Burlet, F. Bourdarot, S. Quezel, J. Rossat-Mignod, P. Lejay, B. Chevalier, and H. Hickey,$\it{J.~Magn.~Magn.~Matter. }~ \textbf{108}$, 202 (1992).

\bibitem{urrs17}T. Yanagisawa,  $\it{Phil. Mag.}~ \textbf{94}$, 3775-3788 (2014).

\bibitem{urrs20} M.N. Wilson, T.J. Williams, Y.-P. Cai, A.M. Hallas, T. Medina, T.J. Munsie, S.C. Cheung, B.A. Frandsen, L. Liu, Y.J. Uemura, and G. M. Luke, $\it{Phys.~Rev.~B}~ \textbf{93}$, 064402 (2016).

\bibitem{urrs21} T.J. Williams, Z. Yamani, N.P. Butch, G.M. Luke, M.B. Maple, and W. J. L. Buyers, $\it{Phys.~Rev.~B}~ \textbf{86}$, 235104 (2012).

\bibitem{urrs22} N.P. Butch and W. J. L. Buyers,  $\it{J. Phys.: Condens. Matter}~ \textbf{22}$, 164204 (2010).

\bibitem{urrs23} S.-H. Baek, M.J. Graf, A.V. Balatsky, E.D. Bauer, J.C. Cooley, J.L. Smith, and N.J. Curro $\it{Phys.~Rev.~B}~ \textbf{81}$, 132404 (2010).

\bibitem{urrs24}P. Das, N. Kanchanavatee, J. S. Helton, K. Huang, R. E. Baumbach, E. D. Bauer, B. D. White, V. W. Burnett, M. B. Maple, J. W. Lynn, and M. Janoschek, $\it{Phys.~Rev.~B}~ \textbf{91}$, 085122 (2015)

\bibitem{urrs25} N. Kanchanavatee, B. D. White,V.W. Burnett, andM. B.Maple, $\it{Philos. Mag.}~ \textbf{94}$, 3681 (2014).

\bibitem{urrs13} M.Yokoyama, H. Amitsuka, S. Itoh, I. Kawasaki, K. Tenya, and H. Yoshizawa, $\it{J. Phys. Soc. Jpn.}~ \textbf{73}$, 545 (2004).

\bibitem{urrs36} T. Sakakibara, H. Amitsuka, T. Goto, K. Sugiyama, Y. Miyako,  M. Date, $\it{Physica B}~ \textbf{177}$, 151 (1992).

\bibitem{urrs27} K. Kuwahara, S. Yoshii, H. Nojiri, D. Aoki, W. Knafo, F. Duc, X. Fabreges, G. W. Scheerer, P. Frings, G. L. J. A. Rikken, F. Bourdarot, L. P. Regnault, and J. Flouquet$\it{Phys.~Rev.~Lett.}~ \textbf{110}$, 216406 (2013).

\bibitem{urrs18}M. Yokohama, H. Amitsuka, S. Itoh, I. Kawasaki, K. Tenya and H. Yoshizawa, $\it{J. Phys. Soc. Jpn. }~ \textbf{73}$, 545 (2004).

\bibitem{urrs19}H. Amitsuka, K. Hyomi, T. Nishioka, Y. Miyako and T. Suzuki, $\it{J.~Magn.~Magn.~Matter}~ \textbf{76\&77}$, 168 (1988).

\bibitem{urrs1} S. R. Hall, G. S. D. King, J. M. Stewart, Eds., Xtal 3.4 User’s Manual. University of Australia: Lamb, Perth (1995).

\bibitem{urrs2} V. F. Sears, in: International Tables of Crystallography, vol. C, ed. A.J.C. Wilson (Kluwer, Dordrecht, 1992) p. 383.

\bibitem{urrs15} K.~A. Ross, L. Harriger, Z. Yamani, W.~J.~L. Buyers, J.~D. Garrett, A.~A. Menovsky, J.~A. Mydosh, and C.~L. Broholm, $\it{Phys.~Rev.~B}~ \textbf{89}$, 155122 (2014).

\bibitem{urrs28} G. Cordier, E. Czech, H. Schaefer, P.; Woll, $\it{J. Less. Comm. Met. }~ \textbf{110}$, 327 (1985).

\bibitem{urrs29} A.A. Menovsky, A.C. Moleman, C.E. Snel, T.J. Gortenmulder, H.J. Tan, T.T.M. Palstra, $\it{J. Cryst. Growth }~ \textbf{ 79}$,  316 (1986).

\bibitem{urrs30} T.E. Mason, B.D. Gaulin, J.D. Garrett, Z. Tun, W.J.L. Buyers, E.D. Isaacs, $\it{Phys. Rev. Lett.}~ \textbf{65}$, 3189 (1990).

\bibitem{urrs31} S. Kawarazaki, Y. Kobashi, T. Taniguchi, Y. Miyako, H. Amitsuka, $\it{J. Phys. Soc. Jpn. }~ \textbf{63}$, 716 (1994).

\bibitem{urrs16} N. H. Andersen and H. Smith, $\it{Phys.~Rev.~B}~\bf{19}$, 384 (1979). 

\bibitem{urrs32} T. T. M. Palstra, A. A. Menovsky, J. A. Mydosh, $\it{Phys.~Rev.~B}~ \textbf{33}$, 6527 (1986).

\bibitem{urrs7}F. Bourdarot, E. Hassinger, S. Raymond, D. Aoki, V. Taufour, L.-P. Regnault and J. Flouquet, $\it{J. Phys. Soc. Jpn. }~ \textbf{79}$, 064719 (2010).

\bibitem{urrs34} T.J. Williams, H. Barath, Z. Yamani, J.A. Rodriguez-Riviera, J.B. Leão, J.D. Garrett, G.M. Luke, W.J.L. Buyers, C. Broholm, unpublished, arXiv:1607.00967 (2016).

\bibitem{urrs35} T.J. Williams, M.N. Wilson, A.A. Aczel, M.B. Stone, G.M. Luke, unpublished arXiv:1607.05672 (2016).


\end{thebibliography}

\end{document}